\documentclass[prl,twocolumn]{revtex4-1}

\usepackage{graphicx}
\usepackage{multirow}
\usepackage{dcolumn}
\usepackage{bm}
\usepackage[normalem]{ulem}
\usepackage{amsmath}
\usepackage{amsfonts}
\usepackage{amssymb}
\usepackage{color}
\usepackage{colordvi}
\usepackage{dcolumn}

\usepackage[colorlinks=true, breaklinks=true, linkcolor=blue,
urlcolor=blue, citecolor=blue]{hyperref}

\begin{document}

\title{First order Fermi surface topological (Lifshitz) transition in an interacting two-dimensional Rashba Fermi liquid}
\author {Yury Sherkunov and Joseph J. Betouras}
\email{J.Betouras@lboro.ac.uk  or Y.Sherkunov@lboro.ac.uk}
\affiliation {Department of Physics and Centre for the Science of Materials, Loughborough University, Loughborough, LE11 3TU, United Kingdom}
\date{\today}

\begin{abstract}

We study the Fermi surface topological transition of the pocket-opening type in a two dimensional Fermi liquid with spin-orbit coupling of Rashba type. We find that the interactions, far from instabilities, drive the transition first order at zero temperature surprisingly in a more pronounced way than in the case of interacting Fermi liquid without spin-orbit coupling. We first gain insight from second order perturbation theory in the self-energy. We then extend the results to stronger interaction, using the self-consistent fluctuation approximation. We discuss existing experimental work on the system BiTeI and suggest further experiments in the light of these results .

 \end{abstract}

\maketitle

\section{Introduction}

The Lifshitz transition \cite{Lifshitz60} (LT) is a Fermi surface topological transition as a result of the change of the Fermi energy and/or band structure. The possibilities that have been usually considered are either a change of the number of Fermi surfaces in a pocket opening or closing transition, or the connection of two parts of Fermi surface in a neck-opening or closing transition. Recently higher order Fermi surface topological or multicritical transitions have been put forward in order to explain unusual properties of correlated materials \cite{Efremov18, Noah19}. The LT can be induced by the variation of an external parameter such as pressure, doping or  magnetic field and has been experimentally observed in many systems such as heavy fermions \cite{Daou06, Bercx12, AokiCelrln16}, iron-based superconductors \cite{Liu10, Xu13, Khan14}, cuprate-based high-temperature superconductors   \cite{Benhabib15, Wu18, Bragan18},  or other strongly correlated electrons system such as layered material $Na_xCoO_2$ \cite{Okamoto10, Slizovskiy15}. It has been also shown to be responsible for the re-entrant superconductivity \cite{Sherkunov18} in uranium-based ferromagnetic superconductors such as $URhGe$ \cite{HuxleyLifshitz11}. The interaction-driven LT has also been proposed and observed in ultracold  fermionic systems \cite{Wang12, vanLoon16, Quintanilla09}.

Of particular interest is the LT in systems with strong spin-orbit coupling (SOC), such as the material BiTeI \cite{Ye15, Hamlin14} which has shown quantum magnetotransport signature of a change in the Fermi surface topology, very relevant to the present work. Two other classes of systems that can exhibit the same physics are ultracold gases \cite{Wang12} and two-dimensional electronic systems that can be formed at the interface between insulating oxides, such as  $LaAlO_3$ and $SrTiO_3$ \cite{Joshua12}. The former represents an important tool in the search for Majorana fermions while the latter exhibits  a range of  interesting phenomena such as  ferromagnetism, superconductivity and  unique magnetotransport properties, for which a universal LT is responsible  \cite{Joshua12, Yin19}. 

Theoretically, it has been shown that interactions, in the region of paramagnetic fluctuations, can play an important part in the LT in a two-dimensional Fermi liquid changing the order of the transition from the second for non-interacting systems to the first for interacting Fermi liquids in a pocket-opening transition \cite{Slizovskiy14}. 
Here we investigate how interactions, outside any phase formation, affect the LT in a two-dimensional Fermi liquid with SOC. In particular, we consider an LT occurring when the bottom of $s=-1$ Rashba subband crosses the level of chemical potential as a result of  variations in doping, as shown in Fig. \ref{Fig1}. We show that, similarly to \cite{Slizovskiy14}, at zero temperature the LT changes its order from the second to the first due to strong paramagnetic fluctuations manifesting themselves as the non-analyticity of the self-energy in the Fermi momenta, $p_F$, calculated from the  bottom of the $s=-1$ Rashba subband (see Fig. \ref{Fig1}). As we show, using the second order of perturbation theory, the self-energy correction takes the form $\Sigma_2\propto p_F \log p_F$ in the presence of SOC while in the absence of it the correction is $\Sigma_2\propto p_F^2 \log p_F$ \cite{Slizovskiy14}. This fact results in an even stronger first order character of the transition in the two-dimensional FL in the presence of Rashba type SOC compared to the absence of it.

In the next sections, we introduce the model and then we gain insight by working in second order perturbation theory. This is followed by summation of diagrams in random phase approximation (RPA) and finally we discuss the wider implications and propose possible experimental directions.

\section{Model}
\begin{figure}[h]
\includegraphics[width=0.4\textwidth]{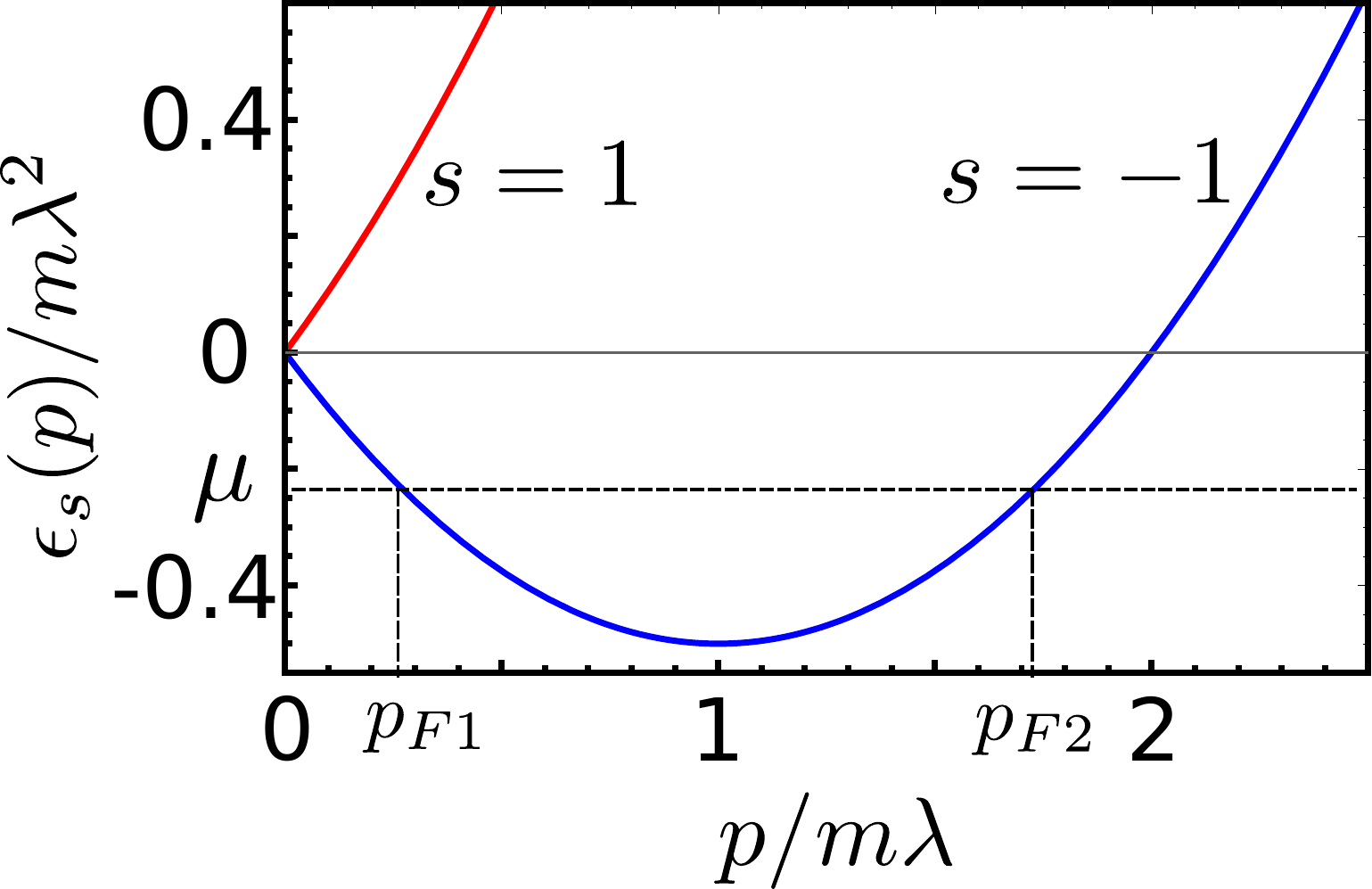}
\caption{Dispersion relation for two helical branches ($s=\pm 1$) of a non-interacting system with Rashba SOC. }
\label{Fig1}\end{figure}

We consider a 2D Fermi liquid with strong Rashba-type SOC  described by the Hamiltonian:
\begin{eqnarray}
H=H_{0}+H_{int}.
\end{eqnarray}
The noninteracting part $H_0$ 
\begin{eqnarray}
H_0=\Sigma_p b^{\dagger}_p[(\frac{p^2}{2m}-\mu)\sigma_0+\lambda (\sigma_x p_y-\sigma_yp_x)]b_p, \label{H0}
\end{eqnarray} 
where $b_p=(b_{p\uparrow},b_{p\downarrow})^T$ is the annihilation operators of a particle,   $\mu$ is the chemical potential, $\sigma_0$ is the $2\times 2$ unit matrix, $\sigma_x$ and $\sigma_y$ are Pauli matrices, and $\lambda$ is the SOC constant, can be diagonalized in the helicity basis 
\begin{eqnarray}
|\mathbf p,s\rangle=\frac{1}{\sqrt 2}(1,ise^{-i\phi(\mathbf p)})^T,\label{basis}
\end{eqnarray}
where $\phi(\mathbf p)=\arctan(p_y/p_x)$ is the angle between the $x-$ axis and momentum $\mathbf p$, and $s=\pm 1$ is the helicity. The dispersion relations for the two helical branches 
\begin{eqnarray}
\epsilon_s(p)=\frac{p^2}{2m}+\lambda s|\mathbf p|\label{disp}
\end{eqnarray}
is shown in Fig. \ref{Fig1}. In this work, we consider the case  $\mu<0$, where the Fermi surfaces are in the branch with $s=-1$.  The most interesting situation will correspond to $\mu\rightarrow \mu_0=-m\lambda^2/2$, where the density of states is divergent.  

We take into account short-ranged interactions of Hubbard type in the system, which read:
\begin{eqnarray}
H_{int}=U\sum_{\sigma\neq\sigma'}\sum_{\mathbf p,\mathbf k, \mathbf q}b^{\dagger}_{k+q,\sigma}b^{\dagger}_{p-q,\sigma'}b_{p,\sigma'}b_{k,\sigma},
\end{eqnarray} 
where $\sigma,\sigma'=\{\uparrow,\downarrow\}$ are the spin indices. In the helicity basis Eq. (\ref{basis}), $H_{int}$ takes the form:
\begin{eqnarray}
H_{int}&=&\sum_{s,s'r,r'}\sum_{\mathbf p,\mathbf k, \mathbf q}U_{r,s}^{r',s'}(\mathbf p,\mathbf k,\mathbf q)a^{\dagger}_{k+q,s'}a^{\dagger}_{p-q,r'}a_{p,r}a_{k,s},\label{Hint}\\
& &U_{r,s}^{r',s'}(\mathbf p,\mathbf k,\mathbf q)=\frac{U}{4}\left(r r' e^{i[\phi(\mathbf p)-\phi(\mathbf p-\mathbf q)]}\right.\nonumber\\
&+&\left. s s' e^{i[\phi(\mathbf k)-\phi(\mathbf k+\mathbf q)]}\right),\label{U}
\end{eqnarray} 
where $a_{p,s}$ is the annihilation operator of a particle in the state  $|\mathbf p,s\rangle$.

\section{Perturbation theory}

We start by analysing the lowest orders of perturbation theory valid for small $U$. Using Matsubara Green's functions, the self-energy correction at first order perturbation theory and at zero temperature is given by the Hartree diagram (the Fock contribution is zero), which leads to $\Sigma_1(p_F,i\omega=0)=Un/2$, where $n$ is the electron density. This contribution is absorbed into $\mu$. The second order contribution to the self-energy is
\begin{eqnarray}
& &\Sigma_2(p_F,i\omega=0,s=-1)=\frac{U^2}{2}\int_q\sum_{s'}\{\chi^0_{00}(q)\nonumber\\
&\times& [1+ss'\cos[\phi(\mathbf p)-\phi(\mathbf p-\mathbf q)]]\nonumber\\
&+&\chi^0_{zz}(q)[1-ss'\cos[\phi(\mathbf p)-\phi(\mathbf p-\mathbf q)]]\}g^0_{s'}(p-q),\label{SOPT}
\end{eqnarray}
  where $q=\{\mathbf q,i\omega\}$ is a four-dimensional momentum, $\int_q\equiv\int \frac{d^2qd\omega}{(2\pi)^3}$, $g^0_{s}(p)=(i\omega-\epsilon_s(p)+\mu)^{-1}$ is the electron Green's function in the helicity basis. The charge, $\chi^0_{00}$ and spin, $\chi^0_{zz}$ susceptibilities of non-interacting particles are defined as the Fourier transform of
  \begin{eqnarray}
  \chi^0_{ij}(\mathbf r,\mathbf r')=\nonumber\\
  -\int_0^\infty d\tau\langle T_\tau\psi^{\dagger}(\mathbf r,\tau)\sigma_i\psi(\mathbf r,\tau)\psi^{\dagger}(\mathbf r',0)\sigma_j\psi(\mathbf r',0)\rangle,\label{suscepdef}
  \end{eqnarray}
with $\psi(\mathbf r,\tau)=\sum_{\mathbf p}b_pe^{i\mathbf p\cdot \mathbf r}$, which can be calculated as \cite{MatiMaslovPRB15}
\begin{eqnarray}
  \chi^0_{ij}(q)=-Tr\int_p\sigma_iG(p)\sigma_jG(p+q)=\nonumber\\
  -\frac{1}{2}\int_pg_r(p)g_s(p+q)F^{ij}_{sr}(\mathbf p+\mathbf q,\mathbf p),\label{Suscept}
 \end{eqnarray}
  where $G(p)=\sum_s\Omega_s(\mathbf p)g^0_s(p)$ is the Green's function in the spin basis, with $\Omega_s(\mathbf p)=\frac{1}{2}[\sigma_0+s(\sigma_x\sin\phi(\mathbf p)-\sigma_y\cos\phi(\mathbf p))]$ and $F^{ij}_{sr}(\mathbf p,\mathbf k)=2\langle p,s|\sigma_i|k,r\rangle\langle k,r|\sigma_j|p,s\rangle$ is the overlap factor. %(check)
  
\begin{figure}[h]
\includegraphics[width=0.4\textwidth]{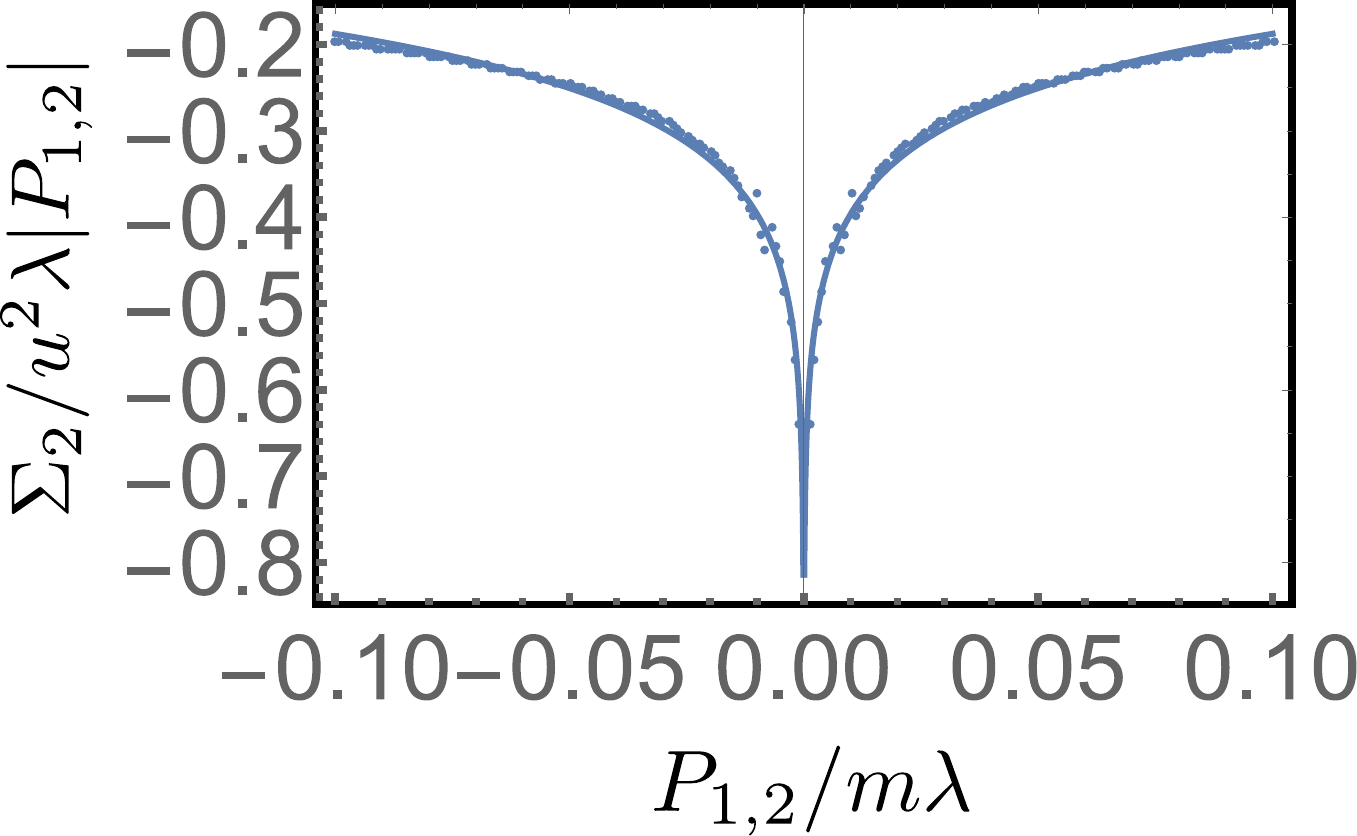}
\caption{Fitting of the self-energy calculated numerically in the second order perturbation theory by $\Sigma_2(P_{1,2}\rightarrow 0)/(|P_{1,2}|u^2\lambda)=a+b\log|\Lambda/P_{1,2}|$, where $\Lambda=10^2m\lambda$,  $a=2.1\times 10^{-2}$ and $b=9.1\times 10^{-2}$ in the vicinity of the bottom of the $s=-1$ band }
\label{Fig1p}\end{figure}

In the vicinity of the bottom of $s=-1$ band, where $\mu\rightarrow\mu_0=-m\lambda^2/2$ and at large transferred momenta  $|\mathbf q|\gg m\lambda\gg|p_{F1,2}-m\lambda|$, the susceptibilities can be estimated as 
 \begin{eqnarray}
 \chi^0_{00}(q)=\chi^0_{zz}(q)=\frac{m}{\pi}\frac{q^2(p_{F2}^2-p_{F1}^2)}{q^4+4m^2\omega^2},\label{chi2order}
 \end{eqnarray}
 which with logarithmic accuracy in the vicinity $p_{F1}\rightarrow m\lambda$ and $p_{F2}\rightarrow m\lambda$   yields
 \begin{eqnarray}
 \Sigma_2(p_{F1,2},i\omega=0,s=-1)\approx -\frac{u^2\lambda |P_{1,2}|}{2\pi^2}\log\left|\frac{\Lambda}{ P_{1,2}}\right|,\label{Sig2}
 \end{eqnarray}
 where $u$ is the Hubbard coupling constant in units of $m\lambda^2$, $P_1=m\lambda-p_{F1}$ and $P_2=p_{F2}-m\lambda$ are the Fermi momenta calculated from the bottom of the $s=-1$ band, and $\Lambda$ is the short wave-length cutoff. We checked the validity of Eq. (\ref{Sig2}) by fitting the result of numerical calculations (see Fig.\ref{Fig1p}). 
 
Eq. (\ref{Sig2}) is similar to the self energy obtained for the system without SOC \cite{Slizovskiy14}, however, the main important difference is that $\Sigma$ in Eq. (\ref{Sig2}) is proportional to the first power of the Fermi momenta $P_{1,2}$, while for the systems without SOC, it is proportional to the second power of the Fermi momentum.  
    
%Similarly to \cite{Slizovskiy14}, 
For the equation of the chemical potential $\mu\sim\mu_0<0$ we find:
\begin{eqnarray}
\mu=\epsilon_{-1}(p_{F1,2})+\Sigma_2(p_{F1,2},i\omega=0,s=-1)\approx \mu_0\nonumber\\
+\frac{P^2_{1,2}}{2m}-\frac{u^2\lambda |P_{1,2}|}{2\pi^2}\log\left|\frac{\Lambda}{ P_{1,2}}\right|,\label{mueq}
\end{eqnarray}
thus, for small $|P_{1,2}|$ the third term becomes large compared to the second one giving rise to a first-order LT. Note that because the third term is proportional to $|P_{12}|$ and the second one is quadratic in $|P_{12}|$, the effect could be more pronounced than in the case of no SOC, where both terms are quadratic in the Fermi momentum \cite{Slizovskiy14}. 

\section{Random phase approximation}

\begin{figure}[h]
\includegraphics[width=0.45\textwidth]{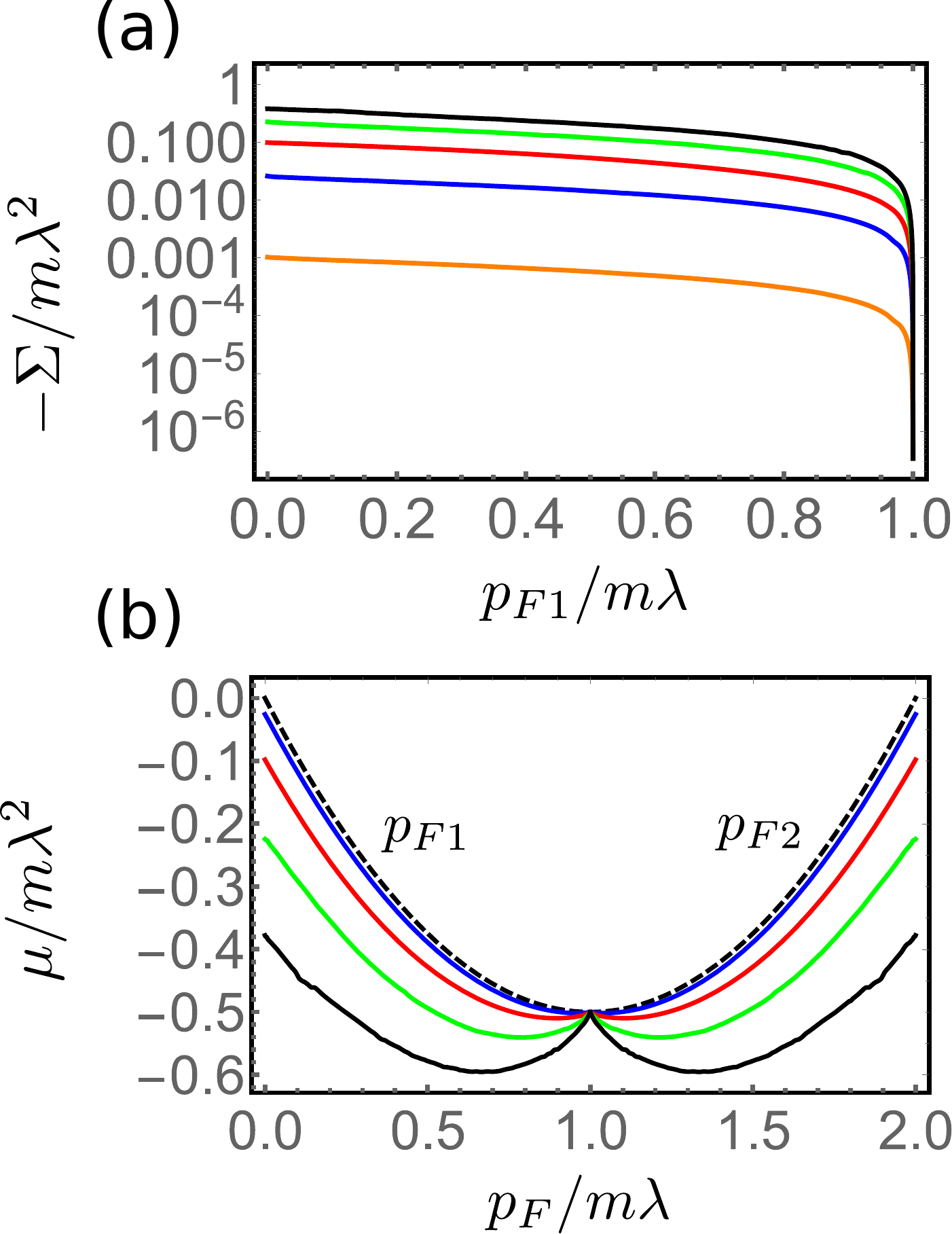}
\caption{ $\Sigma$ as a function of $p_{F1}$ (left)   for $u=0.1$(orange), $u=0.5$ (blue), $u=1$ (red), $u=1.5$ (green), and $u=2$ (black) and $\mu$ as a function of $p_{F1,2}$ (right) for  $u=0.5$ (blue), $u=1$ (red), $u=1.5$ (green), and $u=2$ (black). The dashed curve represents the chemical potential of non-interacting Fermi gas.  }
\label{Fig2}\end{figure}

We now consider larger interaction strengths, but below any Stoner-like instability. Then the effective interaction can be obtained by summing up ring and ladder diagrams \cite{Slizovskiy14}, leading to 
\begin{eqnarray}
\Sigma(p_F,i\omega&=&0,s=-1)=\Sigma_{ring}+\Sigma_{lad},\label{SigRPA}\\
\Sigma_{ring}&=&\frac{u^2}{2}\int_q\sum_r\left[\frac{\chi^0_{00}(q)}{1+u\chi^0_{00}(q)}F_{sr}^{00}(\mathbf p-\mathbf q,\mathbf p)\right.\nonumber\\
&+&\left.\frac{\chi^0_{zz}(q)}{1-u\chi^0_{zz}(q)}F_{sr}^{zz}(\mathbf p-\mathbf q,\mathbf p)\right]g_r(p-q),\label{ring}\\
\Sigma_{lad}&=&\frac{u^3}{2}\int_q\sum_r\left[\frac{[\chi^0_{0y}(q)]^2}{1-u\chi^0_{yy}(q)}F_{sr}^{00}(\mathbf p-\mathbf q,\mathbf p)\right.\nonumber\\
&-&\left.\frac{[\chi^0_{xz}(q)]^2}{1-u\chi^0_{xx}(q)}F_{sr}^{zz}(\mathbf p-\mathbf q,\mathbf p)\right]g_r(p-q).\label{lad}
\end{eqnarray}
Here we put $m=1$ and $\lambda=1$, so all energies and momenta are measured in the unites of $m\lambda^2$ and $m\lambda$. The spin-charge susceptibility matrix is given by Eq. (\ref{Suscept}), with the overlap matrix elements given by
\begin{eqnarray}
F^{00}&=&1+rs\cos[\phi(\mathbf p)-\phi(\mathbf p+\mathbf q)],\nonumber\\
F^{zz}&=&1-rs\cos[\phi(\mathbf p)-\phi(\mathbf p+\mathbf q)],\nonumber\\
F^{0y}&=&-(r\cos[\phi(\mathbf p)]+s\cos[\phi(\mathbf p+\mathbf q)]),\nonumber\\
F^{yy}&=&1+rs\cos[\phi(\mathbf p)+\phi(\mathbf p+\mathbf q)],\nonumber\\
F^{xx}&=&1-rs\cos[\phi(\mathbf p)+\phi(\mathbf p+\mathbf q)],\nonumber\\
F^{xz}&=&-i(r\cos[\phi(\mathbf p)]-s\cos[\phi(\mathbf p+\mathbf q)]).
\end{eqnarray}
The results of numerical calculations of $\Sigma$ and $\mu$ using eq. (\ref{SigRPA}) are presented in Fig.\ref{Fig2}. 

A 2D Fermi liquid with Rashba SOC can support four collective modes: one plasmon mode and three chiral spin modes manifested by poles of charge and spin susceptibilities, which coincides with the solutions of the following equations \cite{MatiMaslovPRB15}:
\begin{eqnarray}
1+u\chi^0_{00}=0 \label{plasmons}
\end{eqnarray}
for plasmons and 
\begin{eqnarray}
1-u\chi^0_{jj}=0,\;\;j=x,y,z, \label{snins}
\end{eqnarray}
for spin collective modes, giving rise to instabilities in Eq. (\ref{ring}) and (\ref{lad}). However, the charge plasmon instability manifests itself only for $\mu>0$ \cite{MatiMaslovPRB15}. Indeed, direct calculation of $\chi^0_{00}$ at $q=0$ and $\omega=0$ for $\mu<0$ leads to $\chi^0_{00}(\mathbf q=0,i\omega=0)=\frac{1}{2\pi}\frac{p_{F1}+p_{F2}}{\sqrt{1+2\mu}}>0$, and, thus, Eq. (\ref{plasmons}) has no real solutions. Note, that the divergence of $\chi^0_{00}$ at the bottom of the $s=-1$ band is due to the divergence of the density of states at this point. The spin susceptibilities $\chi^{0}_{jj}$ at $\mathbf q=0$ are given by\cite{MatiMaslovPRB15}
$\chi^0_{zz}(\mathbf q=0,i\omega)=\frac{1}{4\pi}[(p_{F2}-p_{F1})/2+\omega(\arctan\frac{2p_{F1}}{\omega}-\arctan\frac{2p_{F2}}{\omega})]$ and $\chi^0_{xx}(\mathbf q=0,i\omega)=\chi^0_{yy}(\mathbf q=0,i\omega)=\chi^0_{zz}(\mathbf q=0,i\omega)/2$, which take their maximum values $\chi_{zz}=1/\pi$ and $\chi_{xx}=\chi_{yy}=1/2\pi$  at $\omega=0$ and $p_{F1}=0$, leading to chiral-spin instabilities in Eqs. (\ref{ring}) and (\ref{lad}) at $u_c=\pi$.  Thus, in our analysis we consider $u<u_c$.

For $\mu<\mu_0=-m\lambda^2/2$, the $s=-1$ band of non-interacting fermions is empty, however, in the presence of interactions, the effective energy of fermions bents down leading to opening a pocket. In this case,  Eq. (\ref{mueq}) has four solutions for $p_F$.

\begin{figure}[h]
\includegraphics[width=0.4\textwidth]{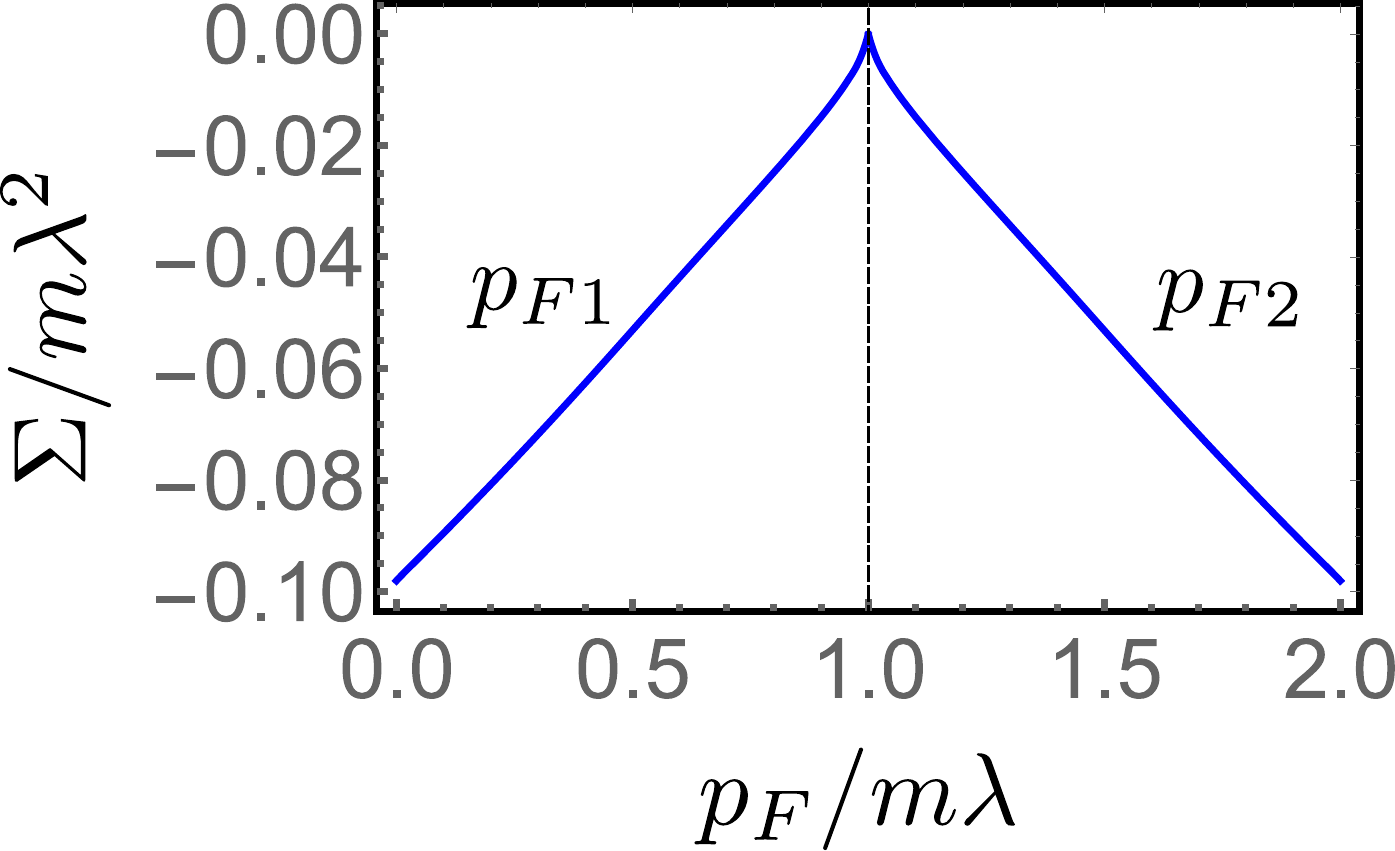}
\caption{ The self-energy $\Sigma$ as a function of $p_F$, for $u=1$. }
\label{Fig3}\end{figure}

In Fig. \ref{Fig3} the different values self-energy corrections at the two values of the Fermi momentum $p_F$ is demonstrated, for completeness.

\begin{figure}[t]
\includegraphics[width=0.4\textwidth]{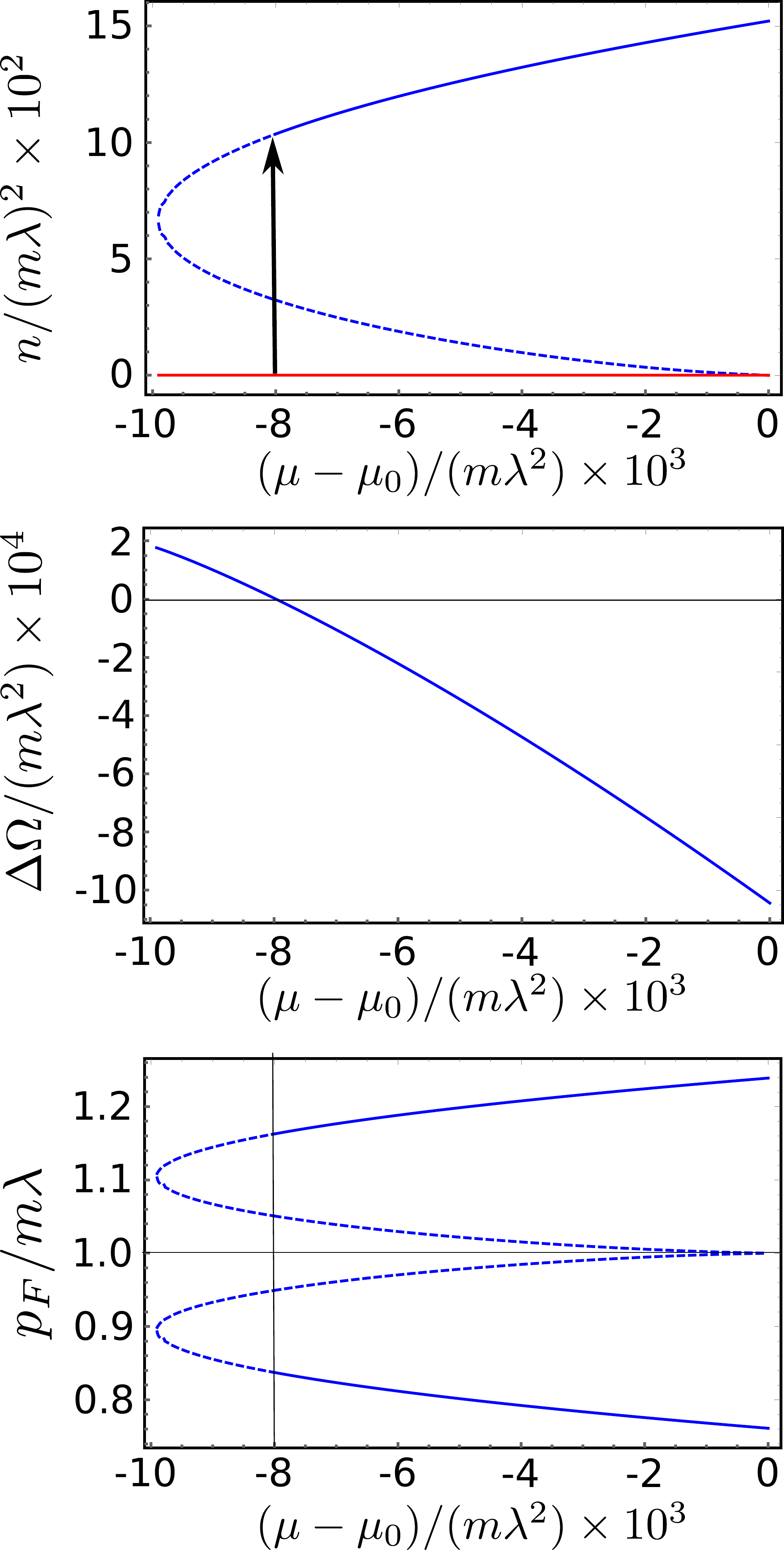}
\caption{Top: electron density $n$ as a function of chemical potential for $u=1$. The red line indicates the states with no pocket for $\mu<\mu_0$, the dashed line corresponds to unstable solutions of Eq. (\ref{mueq}), with arrow showing the phase transition. Middle: potential $\Delta\Omega$ as a function of chemical potential. The stable solutions correspond to negative $\Delta\Omega$. Bottom: Size of the pocket. The dashed curves correspond to unstable solutions. }
\label{Fig4}\end{figure}

In order to estimate which solutions are stable, we 
%follow Ref \cite{Slizovskiy14} and 
estimate the potential $\Omega$ integrating  $d\Omega=-nd\mu$ from the point where the phases with the pocket and no pocket merge. The density of states are given by the Luttinger theorem, $n=\frac{1}{2\pi}(p_{F2}^2-p_{F1}^2)$, which is respected. The results suggesting that in a Fermi liquid with SOC one can expect a first order phase transition in the vicinity of the bottom of the $s=-1$ band are shown in Fig. \ref{Fig4}.

\section{Discussion}

We have showed that the LT of a pocket appearing type in 2D fermions with SOC and in the presence of interactions is discontinuous. This first order transition is more pronounced than in the case without SOC due to the different form of the kinetic energy which acquires a linear in momentum term. The reason of the first order behavior is the competition between kinetic and self-energy contribution to the energy.

This knowledge is needed to disentangle the contribution of different processes that may occur simultaneously and, therefore, to explain behavior that may be attributed to quantum criticality. It is also crucial to realise that unconventional behavior such as temperature dependence of specific heat $C \propto T ln(1/T)$ can be also realised in systems with a LT of pocket appearing/disappearing type \cite{Slizovskiy15}. 

A prime candidate to show this behavior is the giant Rashba semiconductor BiTeI which exhibits a change in the Fermi surface topology in quantum magnetotransport experiments, upon systematic tuning of the Fermi level E$_F$ \cite{Ye15}.  BiTeI has a conduction band bottom that is split into two sub-bands due to the strong Rashba coupling, resulting in a Dirac point. A marked increase (or decrease) in electrical resistivity is observed when E$_F$ is tuned above (or below) this Dirac node, beyond the quantum limit. The origin of this behavior has been shown convincingly to be the Fermi surface topology and essentially it reflects the electron distribution on low-index  Landau levels.  The finite bulk $k_z$ dispersion along the $c$ axis and strong Rashba spin-orbit coupling strength of the system enable this measurement. The Dirac node is independently identified by Shubnikov-de Haas oscillations as a vanishing Fermi surface cross section at $k_z=0$. 
Further measurements of Shubnikov-de Haas oscillations on BiTeI under applied pressures supported the same physics. One high frequency oscillation at all pressures and one low frequency oscillation that emerges between  $0.3$ and $0.7$ GPa has been observed \cite{Hamlin14}, indicating the appearance of a second small Fermi surface.  The suggested explanation is that the chemical potential starts below the Dirac point in the conduction band at ambient pressure and crosses it as pressure is increased. As a result, the pressure brings the system closer to the predicted topological quantum phase transition. 

Regarding experimental observation, there are two promising routes. One is related to recent advances in creating complex oxide heterostructures, with interfaces formed between two different transition metal oxides \cite{Joshua12, Sulpizio14}, which enables the investigation of new physical phenomena in experimentally controlled systems. There is a universal LT demonstrated already in the prototypical LaAlO$_3$/SrTiO$_3$ interface \cite{Joshua12}, between d-orbitals at the core of the observed transport phenomena in this system. At the LT and the critical electronic density, the transport switches from single to multiple carriers. Although the order of the LT it was beyond the scope of the experiment, there was observed a hysteretic behavior as a function of the applied magnetic field near its value at the metamagnetic transition \cite{Ilani19}. As a result, more measurements to clarify our predictions are necessary. The second experimental direction is the ultracold atoms where 2D SO coupled atoms in optical lattices can be realised \cite{Grusdt17}.

In a recent work \cite{Miserev19} a magnetic first order transition has been reported in monolayers of transition metal dichalcogenides with strong SOC, as a result of exchange intervalley scattering. The presence of this interaction would turn our system ferromagnetic.
Finally, it is worth mentioning that there has been growing interest in the effects of SOC on the fermionic Hubbard model in a two-dimensional square lattice. In particular in Ref. \onlinecite{Sun17} it was shown that in the strong coupling limit, the inclusion of SOC leads to the rotated antiferromagnetic Heisenberg model, a new class of quantum spin model. Our work explores another aspect of the same Hamiltonian.

{\it Acknowledgments.} We acknowledge useful discussions and communications with James Hamlin, Shahal Ilani, Jelena Klinovaja,  and Sergey Slizovskiy. The work is supported by EPSRC through the grant EP/P003052/1.

\bibliography{ybs-2.bib}{}
\bibliographystyle{apsrev}
\end{document}